\documentclass[twocolumn,aps]{revtex4}
\usepackage[latin9]{inputenc}
\usepackage{amsmath}
\usepackage{amssymb}
\usepackage{graphicx}
\usepackage{esint}

\makeatletter
\@ifundefined{textcolor}{}
{%
 \definecolor{BLACK}{gray}{0}
 \definecolor{WHITE}{gray}{1}
 \definecolor{RED}{rgb}{1,0,0}
 \definecolor{GREEN}{rgb}{0,1,0}
 \definecolor{BLUE}{rgb}{0,0,1}
 \definecolor{CYAN}{cmyk}{1,0,0,0}
 \definecolor{MAGENTA}{cmyk}{0,1,0,0}
 \definecolor{YELLOW}{cmyk}{0,0,1,0}
}

\usepackage{amsfonts}\usepackage{color}\usepackage{epstopdf}
\DeclareGraphicsExtensions{.pdf,.eps,.png,.jpg,.mps} 

\makeatother

\begin{document}

\title{Experimental Study of Synchronization of Coupled Electrical Self-Oscillators
and Comparison to the Sakaguchi-Kuramoto model}

\author{L. Q. English$^{1}$, Zhuwei Zeng$^{1}$, and David Mertens$^{1}$}

\affiliation{$^{1}$Department of Physics and Astronomy \\
 Dickinson College, Carlisle, Pennsylvania, 17013, USA}

\date{\today}
\begin{abstract}
We explore the collective phase dynamics of Wien-bridge oscillators
coupled resistively. We carefully analyze the behavior of two coupled
oscillators, obtaining a transformation from voltage to effective
phase. From the phase dynamics we show that the coupling can be quantitatively
described by Sakaguchi's modification to the Kuramoto model. We also
examine an ensemble of oscillators whose frequencies are taken from
a flat distribution within a fixed frequency interval. We characterize
in detail the synchronized cluster, its initial formation, as well
as its effect on unsynchronized oscillators, all as a function of
a global coupling strength.
\end{abstract}
\maketitle

\section{Introduction}

The study of spontaneous synchronization is a rich and cross-disciplinary
field of research. The phenomenon can be identified to play a crucial
role in many real biological \cite{buck,walker}, chemical \cite{kiss,fukuda,solomon},
and mechanical systems \cite{bridge,mertens,zhang}, as well as in
neuroscience \cite{michaels,sompolinsky,breakspear}. Powerful mathematical
tools \cite{kura,sakaguchi,kuramoto,strogatz,pikovsky-book,OttAntonsen}
have been developed to analyze synchronization. In spite of the utility
and breadth of the concept, the vast majority of empirical work on
synchronization has been numerical: simple experimental testbeds are
rare in the literature.

A number of complicated experimental systems have been shown to synchronize.
Although some have seen use in testing theories of synchronization,
most are experimentally less accessible. Experiments with pedestrians
on the Millennium Bridge in London \cite{bridge} are difficult to
repeat (for obvious reasons). Lasers \cite{lasers}, Josephson junctions
\cite{josephson}, nickel-electrodes submerged in sulfuric acid \cite{kiss},
and Belousov-Zhabotinsky reactors \cite{bz-reactors} require substantial
experimental setup. Resonantly coupled cell-phone vibrators \cite{mertens}
push the limits of conventional synchronization theory but do not
allow much control in the coupling. Arrays of Belousov-Zhabotinsky
droplets \cite{bz-droplets} only support 1D or 2D lattice topologies.
The circuit of Bergner and coworkers along with that of Gambuzza et
al. offers an implementation of electronic oscillators based on analog
multipliers to model Stuart-Landau oscillators and neurons, respectively,
but even these are complicated circuits \cite{bergner,FHN-circuit}.
In many of these systems, the detailed time- and oscillator-resolved
phase dynamics is difficult to capture experimentally, and the coupling
network topology limited or essentially fixed.

In contrast to all of the given systems, the oscillator circuit of
Temirbayev et al. is simple to build and flexibly couple \cite{rosenblum,moro}.
The underlying circuit design is a Wien-bridge oscillator, an electronic
implementation of a relaxation oscillator. Electronic oscillatory
systems possess a number of advantages for testing theories of synchronization.
They have near-perfect measurement accessibility. They are easy to
prototype or to fabricate in bulk. The coupling between oscillators
can be varied from all-to-all to lattice to complex network by merely
rearranging wires. Wien-bridge oscillators can be made to exhibit
phase-oscillator behavior with near-constant amplitude, even when
coupled to other oscillators. As we will show in this paper, coupled
Wien-bridge oscillators can be quantitatively modeled by the simple
Sakaguchi-Kuramoto model \cite{sakaguchi}.

This paper is structured as follows. In Section II, we describe the
experimental system, detailing individual and coupled circuit designs.
In Section III we present a series of measurements for a pair of coupled
oscillators in which one oscillator's uncoupled speed is increased
with each measurement, and use this data to obtain a simple phase-oscillator
model for our system. In Section IV we explore the synchronization
dynamics of a system of 19 and 20 coupled oscillators in which the
coupling topology is systematically altered, and present measurements
of the transient approach to synchronization.

\section{Experimental system and setup}

\begin{figure}
\includegraphics[width=2.75in]{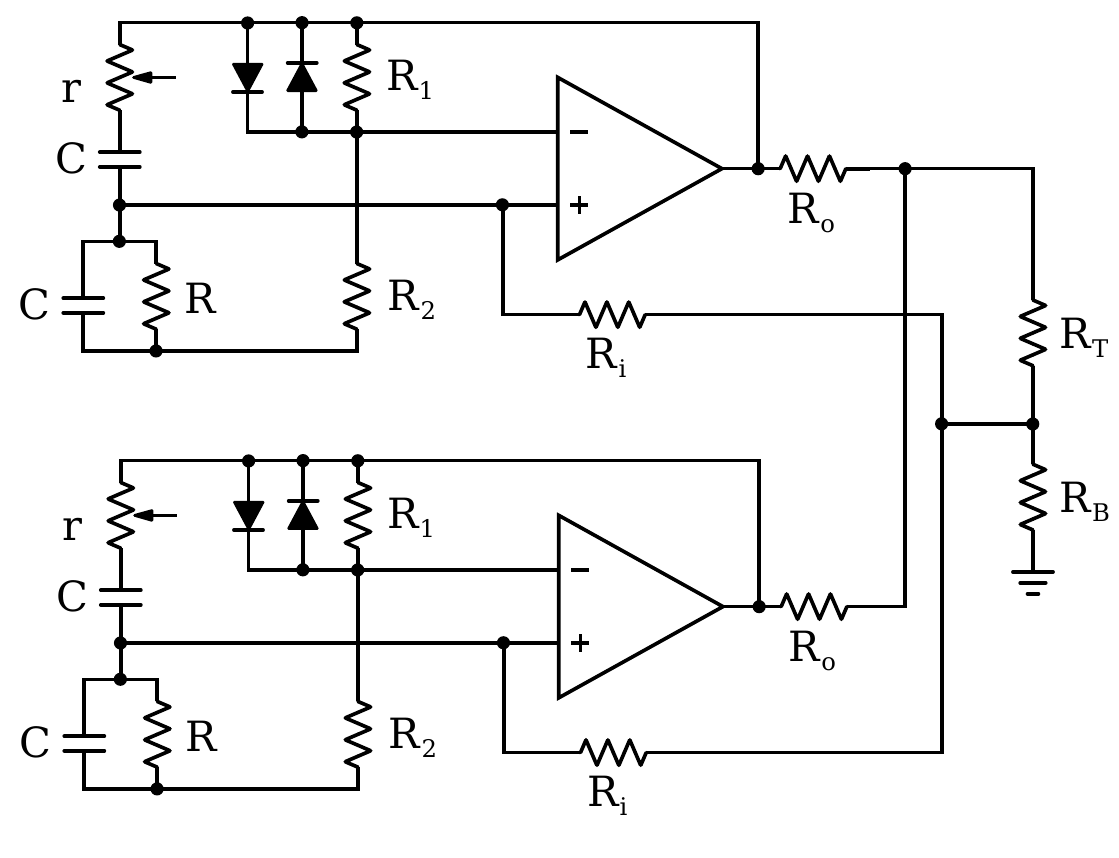} \protect\caption{Circuit diagram of two coupled Wien-bridge oscillators. The intrinsic
frequency is determined by the values of r, R and C. The resistor
$R_{i}$ receives the voltage from the shared voltage divider, whereas
the resistor $R_{o}$ transmits to the voltage divider. For our experiments
we have $R_{1}=27\,\mbox{k\ensuremath{\Omega}}$, $R_{2}=3\,\mbox{k\ensuremath{\Omega}}$,
$C=0.1\,\mbox{\ensuremath{\mu}F}$, $R=4.7\,\mbox{k\ensuremath{\Omega}}$,
$r$ represents a $5\,\mbox{k\ensuremath{\Omega}}$ potentiometer,
$R_{i}=62\,\mbox{k}\Omega$, and $R_{o}=1.1\,\mbox{k}\Omega$.}
\label{2oscsetup} 
\end{figure}

Figure \ref{2oscsetup} gives the circuit diagram of two Wien-bridge
oscillators coupled to one another. We will first consider the circuit
for uncoupled oscillators, obtained by removing the $R_{i}$ and $R_{o}$
resistors. Each oscillator principally consists of two branches, the
resistor/capacitor branch and the resistor/diode branch, configured
as a non-inverting amplifier. The resistor/capacitor branch sets the
primary frequency of the oscillator, $\omega_{0}=\sqrt{\frac{1}{RrC^{2}}}$.
We can easily change this frequency by altering the resistance of
the potentiometer $r$, and we operate our oscillators between $200\,\mbox{Hz}$
and $400\,\mbox{Hz}$. The resistor/diode branch dictates the shape
of the waveform and its sensitivity to perturbations. A pure but sensitive
sine-wave results by setting the gain, $\left(R_{1}+R_{2}\right)/R_{2}$,
to 3. To produce coupled oscillations with robust amplitudes, the
gain for our setup is 10. The signal at such a high gain is far from
sinusoidal, characterized by multiple harmonics, but the constant
amplitude justifies treating them as phase oscillators.

\begin{figure}
\includegraphics[width=3in]{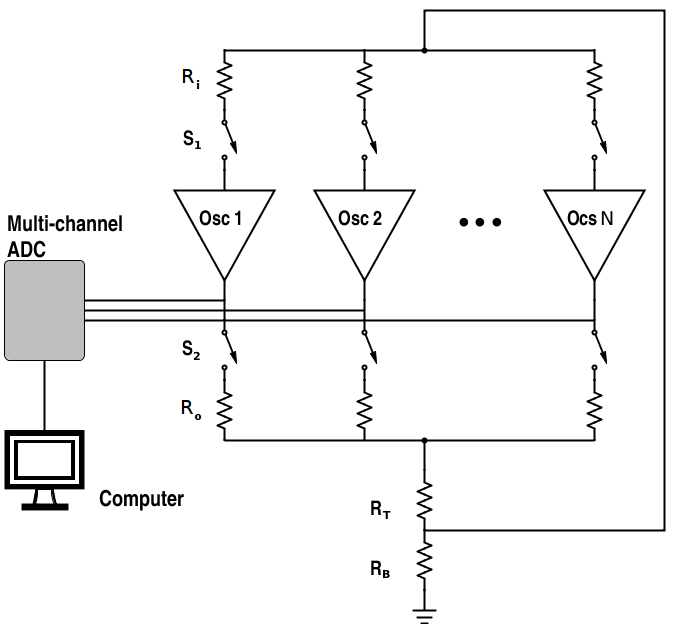} \protect\caption{A schematic of the $N$ Wien-bridge oscillators. We incorporate two
sets of switches, one to prevent particular oscillators from ``talking'',
and the other to prevent them from ``listening'' to the others.
The oscillillators operate at frequencies between $200\,\mbox{Hz}$
and $400\,\mbox{Hz}$, and their voltages are sampled every $0.1\,\mbox{ms}$.}
\label{Nosc} 
\end{figure}

The topology, direction, and nature of the coupling is highly configurable.
In Fig.~\ref{2oscsetup}, two oscillators are coupled via two additional
resistors for each oscillator, namely the $R_{o}$ output resistor
and the $R_{i}$ input resistor \cite{rosenblum}. This coupling method
can then be easily extended to $N$ oscillators, as shown in Fig.~\ref{Nosc}.
Connecting the oscillators to a common tank circuit would produce
an experiment equivalent to the Millennium Bridge \cite{bridge}.
Connecting the oscillators to a tank circuit with two resonant frequencies
would give an analog to the coupled rotors of Mertens and Weaver \cite{mertens}.
Connecting the oscillators in series using unity-gain buffers can
produce a 1D nearest-neighbor lattice suitable for testing the renormalization
theory of Lee et al. \cite{chain-universality}, while purely resistive
coupling would lead to non-local effects suitable for testing theories
of chimeras \cite{chimera-theory}. Complex network coupling could
be used to test the coupling extraction methods of Kraleman et al.
\cite{reconstructing-networks}. Temirbayev et al. introduced additional
electronics that led to a nonlinear shift in the phase coupling \cite{rosenblum}.
In this paper, we focus on a form of directed coupling in which all
oscillators are influenced by a selected subset of the oscillators,
which we call all-to-some coupling.

As shown in Fig. \ref{Nosc}, the individual input resistors are all
connected to a common bus, as are the individual output resistors.
The coupling strength is determined by the settings of the voltage-divider
resistors, $R_{B}$ and $R_{T}$, where we maintain the constraint
$R_{B}+R_{T}=4\,\mbox{k\ensuremath{\Omega}}$. (To see why, notice
that $R_{i}\gg R_{B}$ implies that the voltage drop across $R_{T}+R_{B}$
is the average of the oscillator output voltages. In that case, the
mid-point between $R_{T}$ and $R_{B}$ serves as a simple voltage
divider.) We have also included DIP switches, $S_{1}$ and $S_{2}$,
at every oscillator's input and the output, which serves both as a
measurement convenience and as a simple means for changing which oscillators
can effect and are effected by the common signal on the bus. With
all switches closed, we obtain the conventional all-to-all coupling.
By opening an oscillator's $S_{1}$ and $S_{2}$, we isolate the oscillator
and can measure its uncoupled behavior. By closing $S_{1}$ and $S_{2}$
for an individual oscillator and opening $S_{1}$ and $S_{2}$ for
all other oscillators, we can measure the oscillator's self-coupled
behavior. Closing $S_{1}$ and $S_{2}$ for only two oscillators produces
the pair-coupled configuration shown in Fig. \ref{2oscsetup}. Finally,
by closing $S_{1}$ for all oscillators and selectively opening $S_{2}$,
we couple all oscillators to the common bus but choose which oscillators
contribute to the common bus.

The voltages at the output of each oscillator are simultaneously digitized
using a 14-bit multi-channel DAC with a time resolution of $0.1\,\mbox{ms}$.
The data is streamed to and stored at an acquisition card of a computer.
We trigger the measurement using a pulse generator. For some measurements
where transients are recorded, the pulse generator also drives an
analog switch (ADG 452B). This switch, when closed, bypasses resistor
$R_{B}$ to ground and essentially turns the coupling strength to
zero. Upon opening the switch, the coupling is abruptly turned on.

There are many frequencies that describe the behavior of these oscillators.
In some of our measurements we will focus on the time series of frequencies
averaged over individual oscillation periods, termed \emph{dynamic
frequencies}. To compute the dynamics frequencies, we note the time
of all upward zero crossings of our voltage time series, $T_{1},\,T_{2},\,\dots,\,T_{N}$.
From these zero crossings we can compute the time series $\left(t_{i},\,f_{dyn,i}\right)=\left(\frac{T_{i}+T_{i+1}}{2},\frac{1}{T_{i+1}-T_{i}}\right)$.
The time scales of the interactions among oscillators are much longer
than the oscillators' individual periods, so this gives a suitable
characterization of the instantaneous behavior of each oscillator.
We can also compute the \emph{average frequency} over the whole time
series using $\bar{f}=\frac{N-1}{T_{N}-T_{1}}$. With both $S_{1}$
and $S_{2}$ open, the frequency of an individual oscillator is stable
and is termed the \emph{uncoupled frequency}. This frequency can be
changed by adjusting the oscillator's potentiometer. The frequency
is also stable if both $S_{1}$ and $S_{2}$ are closed but the switches
for all other oscillators are open. This \emph{self-coupled frequency}
can be effected by the oscillator's potentiometer as well as the resistors
on the common bus, $R_{T}$ and $R_{B}$. When multiple oscillators
are coupled together, comparing dynamic frequencies to self-coupled
frequencies is advantageous because some degree of self-coupling operates
in both contexts.

For the ensemble measurements presented in Section IV, the self-coupled
frequencies of the oscillators are set so that they are roughly equally
spaced and cover a range from $240\,\mbox{Hz}$ to $315\,\mbox{Hz}$.

\section{Pairwise Interactions}

Let us begin by examining the dynamics of just two coupled oscillators.
We would expect two such oscillators to come into synchrony if the
coupling strength is sufficiently strong and the natural frequency
mismatch sufficiently small. In order to test this we performed a
series of measurements. The coupling was set as strong as possible
by setting $R_{B}=4\,\mbox{k\ensuremath{\Omega}}$ and $R_{T}=0\,\mbox{\ensuremath{\Omega}}$.
We kept the uncoupled frequency of one oscillator fixed while varying
that of the other oscillator.

\begin{figure}
\includegraphics[width=3.6in]{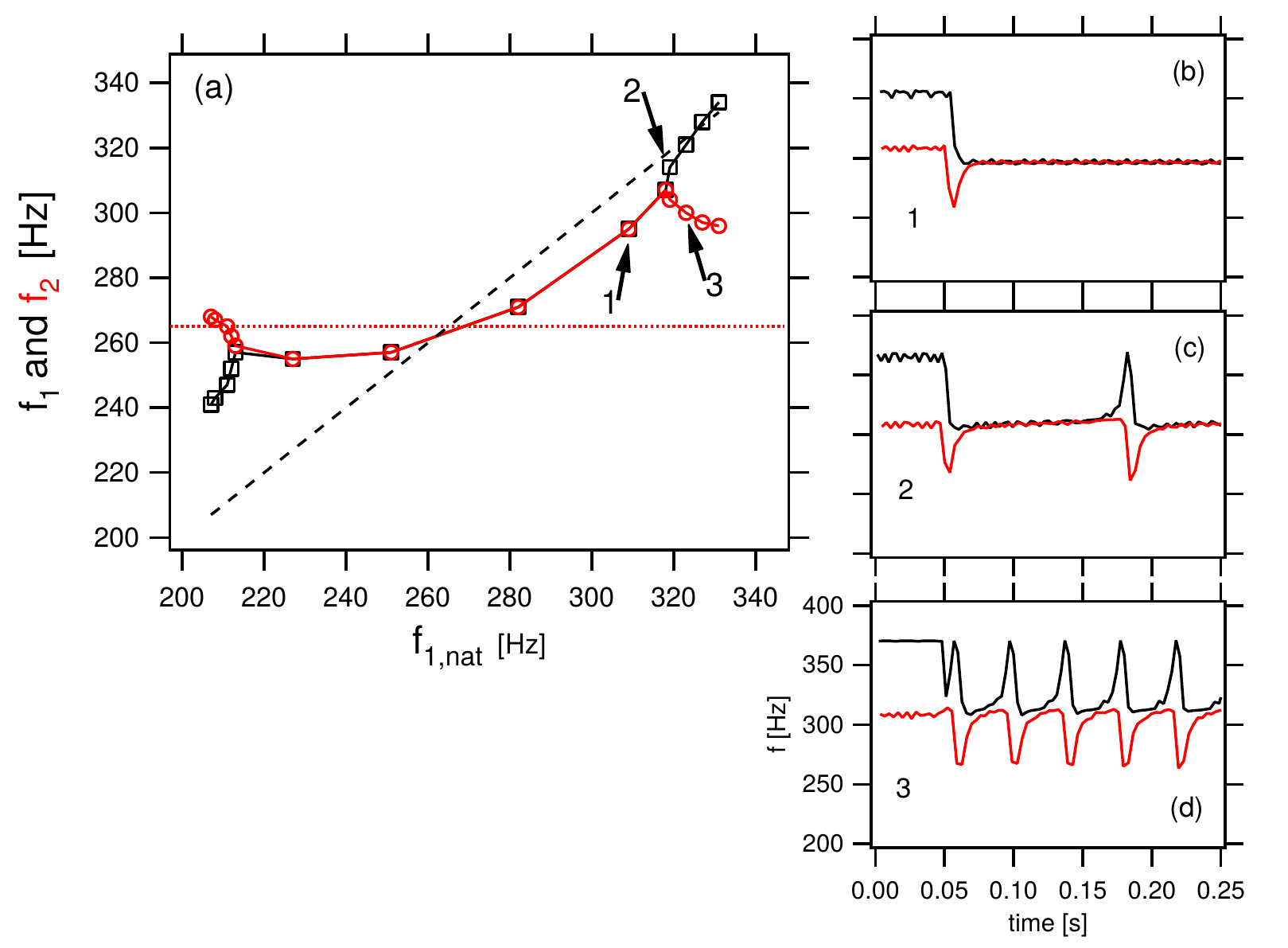} \protect\caption{(Color online) (a) Average oscillator frequencies as we vary one of
the oscillators' natural frequencies, indicating the range of frequency
mismatch over which two oscillators will still synchronize. These
are based on the average period of oscillation over the full $500\,\mbox{ms}$
time series. The dashed black and dotted red lines indicate the self-coupled
frequencies of oscillator 1 and 2, respectively; notice that we vary
the frequency of the first, but not the second. (b-d) The dynamic
frequencies of the two oscillators when coupled, corresponding to
points enumerated in (a). The coupling is initially set to zero and
then abruptly turned on at $t=0.05\,\mbox{s}$. For (b), the two oscillators
quickly come into stable synchronization, but for (c) and (d) this
is not the case, as dephasing occurs in bursts. These bursts occur
more frequently with larger frequency mismatch.}
\label{2oscdata} 
\end{figure}

Figure \ref{2oscdata}(a) summarizes the series of measurements. We
plot the self-coupled frequency of the variable oscillator horizontally,
and the average frequencies of the two oscillators vertically. We
observe that over a large range of frequency settings, the oscillator
pair manages to synchronize to the same frequency. The dashed black
line and the dotted red line show the self-coupled frequencies of
the two oscillators, respectively. Between about $215\,\mbox{Hz}$
and $320\,\mbox{Hz}$, the two oscillators remain phase locked, but
outside of that interval, the frequencies split and tend toward their
self-coupled frequencies.

In order to examine what happens at the branching points more closely,
the right subfigures of Fig.~\ref{2oscdata} depict how the dynamic
frequencies vary over time. The coupling is turned on (via the analog
switch) at time $0.05\,\mbox{s}$. It is clear that when the pair
manages to synchronize, Fig.~\ref{2oscdata}(b), the frequencies
quickly approach each other and remain locked for the entirety of
the run.

Figure \ref{2oscdata}(c) and (d) show how this synchronized state
is lost. In (c), we are just beyond the frequency cutoff for synchronization.
The two dynamic frequencies initially come together and remain close
for many cycles, just like in (b). But then, an abrupt phase slippage
occurs (here around $0.18\,\mbox{s}$), at which time the frequency
of the faster oscillator momentarily speeds up while the slower oscillator
momentarily slows down. During this time the faster oscillator laps
the other, and upon catching up their frequencies again match over
multiple cycles before the process repeats itself. In (d) the phase
breakage occurs more frequently in time. However, even here the frequencies
stay intermittently matched over multiple periods of oscillation.

These data present a number of behaviors that are inconsistent with
the original Kuramoto model. In Fig.~\ref{2oscdata}(a), the synchronized
frequencies exhibit upward curvature; the Kuramoto model predicts
linear behavior. At the tail ends of the unsynchronized behavior,
the frequencies of the faster oscillators are even faster than their
self-coupled frequencies, whereas the Kuramoto model predicts convergence.
In (c) and (d), the oscillators' extreme speeds do not coincide in
time, but instead the fast oscillator spikes slightly ahead of the
slow oscillator. Upon initial inspection, we expected that a complicated
model would be necessary to capture all of these details. Rather than
attempt a theoretical derivation of the coupling behavior, we chose
instead to find a phenomenological phase oscillator model that explains
the system.

To obtain the parameters for our coupling function, we employed an
approach similar to that of Kralemann et al. \cite{reconstruction},
which we describe briefly. Measurements of Wein-bridge oscillators
give their voltage. The voltage time series $V\left(t\right)$ must
be transformed into a phase time series $\theta\left(t\right)$, called
the protophase, by inverting the projection $V\left(t\right)=V_{0}\cos\left(\theta\left(t\right)\right).$
One typically employs the Hilbert transform to obtain the protophase
of each oscillator \cite{pikovsky-book}. If the protophases were
taken as the underlying phases, the abrupt phase velocity changes
inherent in the behavior of the relaxation oscillators would lead
to apparent self-coupling terms. That is, $\dot{\theta}$ depends
upon $\theta$. We wish to transform the protophases such that the
phase velocity extracted from the uncoupled oscillator data is constant
and does not depend upon phase. Therefore, the next step is to obtain
a mapping from protophase $\theta\left(t\right)$ to phase $\phi\left(t\right)$.
Kralemann suggests obtaining the mapping as $\phi\left(\theta\right)=\omega\int_{0}^{\theta}d\theta'\,\sigma\left(\theta'\right),$
where $\sigma\left(\theta\right)$ is the probability density of the
protophase. We instead obtained the mapping $\theta\to\phi$ directly
from the uncoupled time series by assuming that as $\theta$ proceeds
from $-\pi$ to $+\pi$ at a variable rate, $\phi$ proceeds from
$-\pi$ to $+\pi$ at a constant rate. This set of data forms the
basis of a simple interpolation-based mapping that performed as well
as the more sophisticated mapping techniques that we tried.

Our next step is to obtain the parameters for a phase oscillator model
that adequately describes our results. Knowing that the original Kuramoto
model was insufficient, we started by attempting to fit our data to
the Sakaguchi-Kuramoto model \cite{sakaguchi}, which in this case
takes the form 
\begin{equation}
\dot{\phi}_{i}=\omega_{i}+\frac{K}{N}\sum_{j=1}^{N}\sin\left(\phi_{j}-\phi_{i}-\alpha\right).\label{ks-model}
\end{equation}
Here $\omega_{i}$ represents the uncoupled (angular) frequency of
oscillator $i$, which we computed from the average frequency of the
uncoupled oscillator time series. We estimated the $\dot{\phi}_{i}$
by performing numerical differentiation on the $\phi_{i}$ time series.
To obtain $K$ and $\alpha$, we simultaneously fit the data from
both oscillators and across all measurements, obtaining $\alpha=0.509\,\mbox{rad}$
and $K=376\,\mbox{rad/s}$.

\begin{figure}
\includegraphics[width=3.2in]{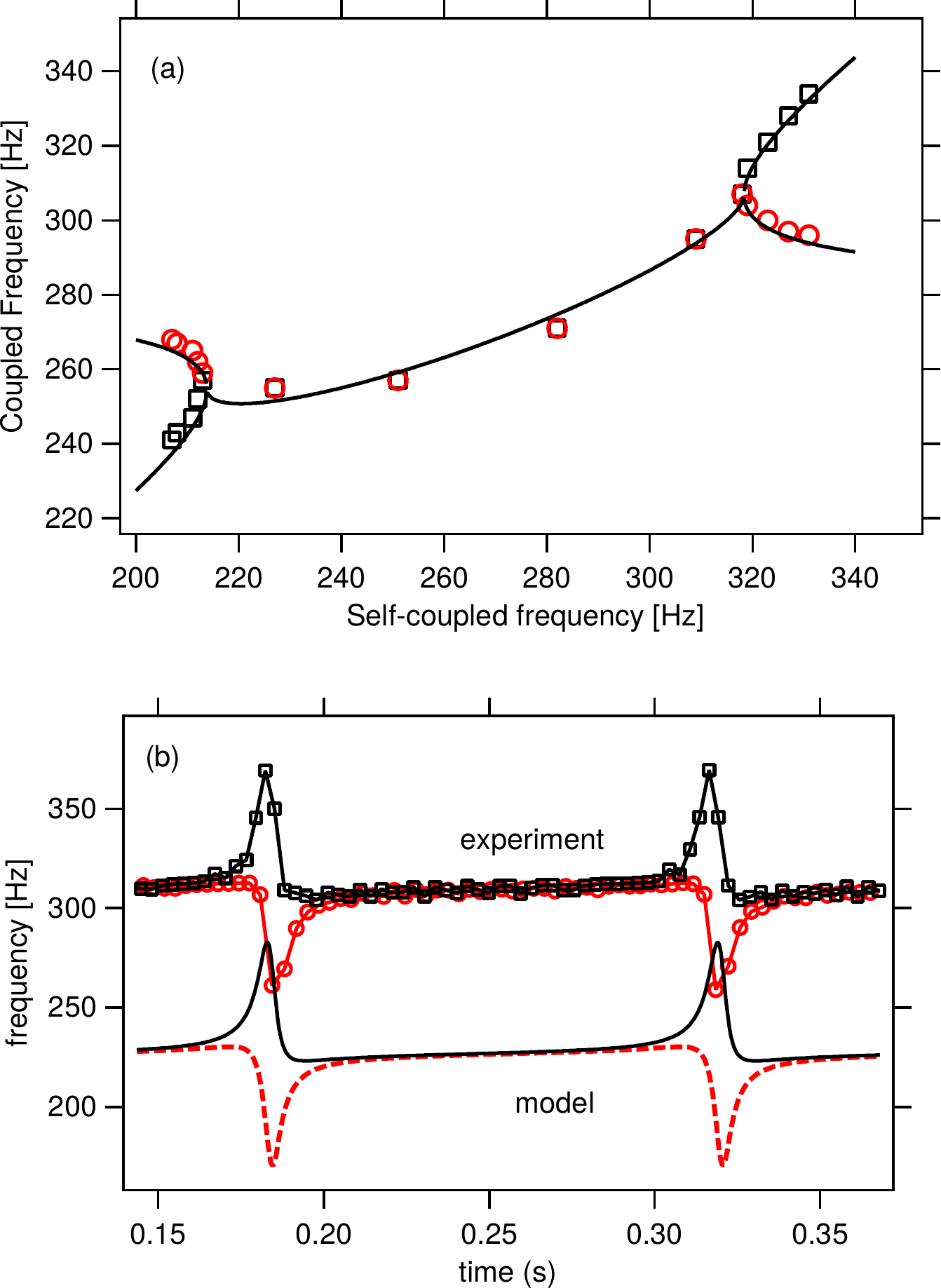} \caption{(Color online) (a) A comparison between the theoretical prediction
of the Sakaguchi-Kuramoto model, Eqs.(\ref{SK_pred_sync}) and (\ref{SK_pred_unsync}),
for the 2-oscillator synchronization-frequency (solid line) and the
experimental data points shown previously (black squares and red circles,
explained in Fig.~\ref{2oscdata}). Good quantitative agreement is
found across the whole set of experiments. (b) Just above the highest
phase-locked frequency, rapid phase-slippage events at regular time
intervals were observed in the experiment: black squares and red circles
with connecting lines represent the measured dynamic frequencies of
the faster and slower oscillator, respectively. The phenomenon is
faithfully reproduced in simulations of the Sakaguchi-Kuramoto model,
for which the black solid line and red dashed line represent the faster
and slower oscillator, respectively. (In the graph, the trace obtained
by numerical simulation is shifted lower for visual clarity.)}
\label{comp} 
\end{figure}

Using these best-fit parameters obtained from the extracted phase
data, we can now ascertain how well the Sakaguchi-Kuramoto model performs
in matching experimental results. For two oscillators, it is fairly
straightforward to derive a prediction for the synchronization frequency
as a function of mismatch in self-coupled frequency. One thing to
remember is that the self-coupled frequency is governed by the case
where $N=1$, coming out to $\omega_{i,self}=\omega_{i}-K\sin\alpha$,
whereas the pair-coupled behavior is governed by the case where $N=2$.
Using this relationship and employing various trigonometric identities,
the prediction for the mutually entrained frequency is obtained as
\begin{align}
f_{entrained}= & \frac{f_{1,self}+f_{2,self}}{2}+\frac{K\sin\alpha}{4\pi}\nonumber \\
 & -\frac{K\sin\alpha}{4\pi}\left(\sqrt{1-4\pi^{2}\frac{\left(f_{1,self}-f_{2,self}\right)^{2}}{K^{2}\cos^{2}\alpha}}\right).\label{SK_pred_sync}
\end{align}
When the oscillators do not mutually entrain, it is possible to compute
the amount of time for the faster oscillator to lap the slower oscillator,
as well as the phase accumulated by each oscillator. Dividing the
phase by the time gives the average phase velocity, from which the
average unsychronized frequencies can be calculated: 
\begin{align}
f_{coupled\pm}= & \frac{f_{1,self}+f_{2,self}}{2}+\frac{K\sin\alpha}{4\pi}\nonumber \\
 & \pm\frac{K\cos\alpha}{4\pi}\left(\sqrt{4\pi^{2}\frac{\left(f_{1,self}-f_{2,self}\right)^{2}}{K^{2}\cos^{2}\alpha}-1}\right).\label{SK_pred_unsync}
\end{align}
The faster oscillator's average frequency is $f_{coupled+}$, while
the slower oscillator's average frequency is $f_{coupled-}$. This
prediction matches the experimental data extremely closely, as shown
in Fig. \ref{comp}(a). The figure superimposes the analytical result
of Eqs. (\ref{SK_pred_sync}) and (\ref{SK_pred_unsync}) on the data
already presented in Fig. \ref{2oscdata}. We reiterate that the solid
line is not a direct fit to the frequency data points, but the theoretical
prediction of the model for our given $\alpha$ and $K$.

The Sakaguchi-Kuramoto model accurately predicts the regime over which
the two Wien-bridge oscillators synchronize. The bifurcation occurs
at the boundary between Eqs. (\ref{SK_pred_sync}) and (\ref{SK_pred_unsync}),
which is $\left|f_{1,self}-f_{2,self}\right|=K\,\cos\alpha/\left(2\,\pi\right)$.
For our parameters, this condition translates into a predicted maximum
self-coupled frequency of $f_{1,self}=319\,\mbox{Hz}$, and a minimum
frequency of $214\,\mbox{Hz}$. Inspection of Fig. \ref{comp} for
these cutoffs reveals excellent agreement with experimental data.

Other features, such as the curvature and asymptotic frequencies,
are also correctly predicted. The entrained frequency is predicted
to have concave-up curvature. Indeed, just above the the lower-frequency
bifurcation point, the mutually entrained frequency \emph{drops} with
increasing $f_{1,self}$, a counterintuitive feature seen in both
the prediction and the data. The frequencies at the tail-ends of the
measurements can be approximated by assuming that the $-1$ term in
Eq. (\ref{SK_pred_unsync}) is negligible, leading to an approximation
$f_{coupled+}\approx f_{self}+K\sin\alpha/\left(4\pi\right)$; since
$\alpha$ is positive, the pair-coupled unsynchronized frequencies
far from the entrainment bifurcation will be larger than the self-coupled
frequencies. 

As we have seen, above the maximum phase-locked frequency the two
oscillators stay in phase with one another for most of the time, but
regular phase-slippage events now occur. The time separation of these
events decreases with larger oscillator frequency difference. This
phenomenon can be reproduced in simulations of the Sakaguchi-Kuramoto
model, as depicted in Fig. \ref{comp}(b). In particular, in both
the experimental data and the numerical simulations, we observe the
oscillator with the larger self-coupled frequency to break away from
the common frequency first, followed in short order by the lower-frequency
oscillator. Notice that the simulations also capture the time duration
of the phase-slippage events as well as the magnitude of deviation
from the common frequency in both directions. This detailed agreement
further demonstrates the direct applicability of the Sakaguchi-Kuramoto
model to this experimental system of electrical oscillators.

The quantitative accuracy of the Sakaguchi-Kuramoto model is surprising
in part because the model is inconsistent with some of the data that
we have not presented here. In particular, the phase time series obtained
from the self-coupled data can be used to obtain phase velocity as
a function of phase, and these phase velocities clearly show a phase
dependence. Theorists and modelers often assume that the coupling
function can be written as $\Gamma\left(\phi_{j}-\phi_{i}\right)$;
Miyazaki and Kinoshita present a concise explanation for why this
should be \cite{bz-reactors}. However, any such model (the Sakaguchi-Kuramoto
model being among them) would predict that the self-coupled speeds
are related to the uncoupled speeds by a constant offset. They are
unable to explain any phase dependence. Fortunately, it appears that
these discrepancies are only minor matters.

Our Wien-bridge oscillators were specified to operate in a regime
that is characterized by a relaxation oscillator. The dynamics of
the protophases seem complicated and aspects of the coupled behavior
seemed counter-intuitive. In spite of these complexities, the underlying
phase dynamics are surprisingly simple, and can be characterized by
the simple and well-studied Sakaguchi-Kuramoto model. Next, we turn
our attention to collective behavior that arises simply by altering
the coupling topology of an ensemble of such oscillators.

\section{Ensemble Measurements}

The asymmetries found in the pairwise data have implications for the
coupled behavior of larger collections of oscillators. To explore
these implications, we prepared a set of 19 oscillators, initially
coupled together in an all-to-all network. The self-coupled frequencies
of the ensemble were roughly evenly spaced over a frequency range
of $\Delta f_{self}=75\,\mbox{Hz}$. In this section we explore the
asymmetries, which become even more striking when we consider an all-to-some
network.

\begin{figure}
\includegraphics[width=3.2in]{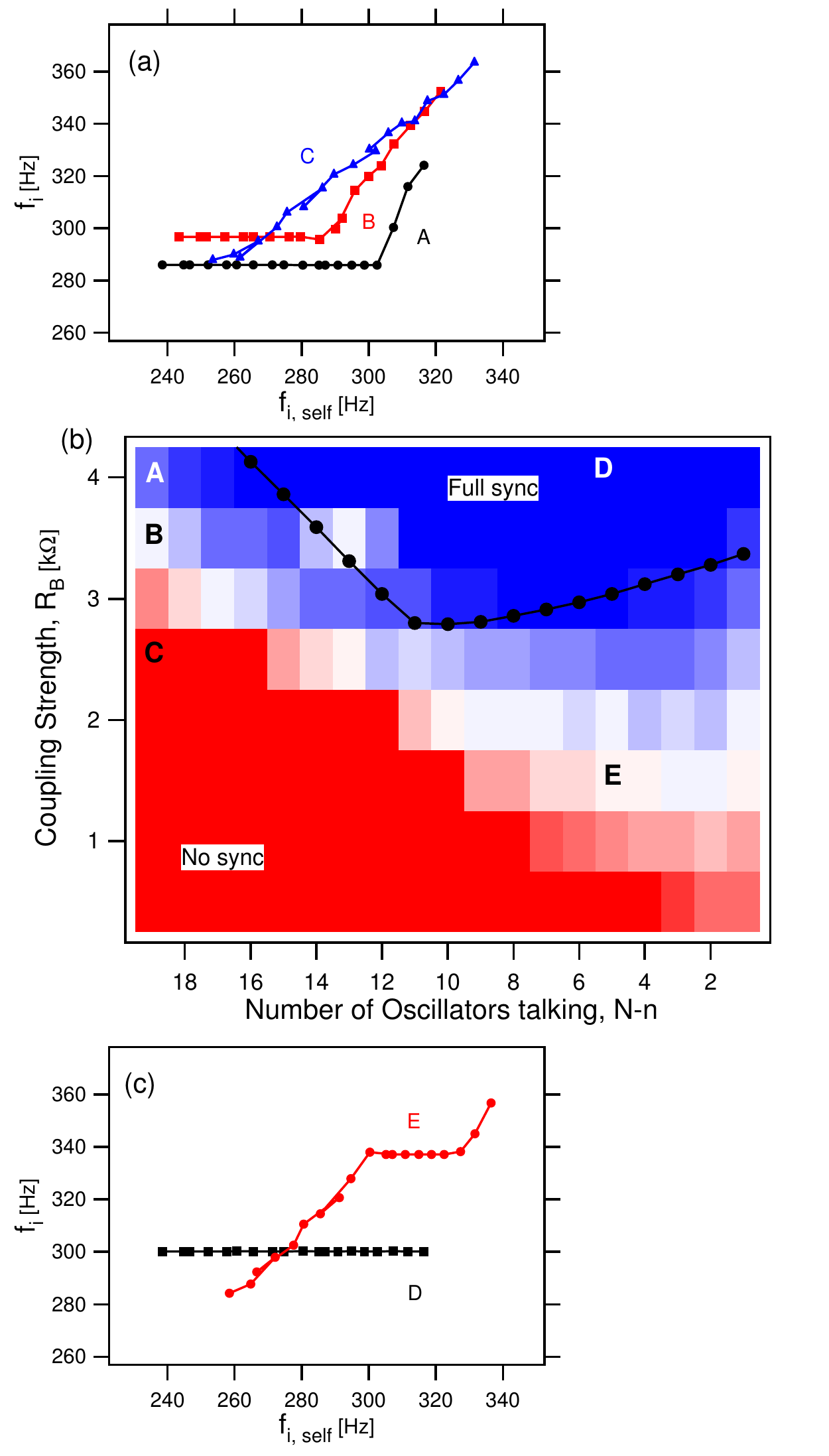}

\protect\caption{(Color online) (a) Plot of long-time average dynamic frequencies vs
self-coupled frequencies for three coupling strengths. The strength
decreases from A to C, as indicated by y-value of the same letter
in the next figure. (b) The y-axis represents the coupling strength
between oscillators as given by the value of $R_{B}$. The x-axis
indicates the number of talkers. As we move from left the right, incrementally
more low-frequency oscillators are silenced (by opening switch $S_{2}$).
The color indicates the size of the largest cluster, ranging from
0 (red, no sync) to 19 (blue, full sync). The line and dots indicates
the theoretically predicted threshold for full synchronization. (c)
Plot similar to (a) for only five talkers. As with (a), the strength
is indicated by the y-value of the same letter in figure (b).}
\label{ps} 
\end{figure}

We begin by systematically examining the fully coupled configuration
over a range of coupling values. Figure \ref{ps}(a) reports the average
dynamic frequency of all oscillators for three different values of
coupling for this configuration. Series A is for the strongest coupling
at $R_{B}=4\,\mbox{k\ensuremath{\Omega}}$ while Series C is for the
weakest coupling at $R_{B}=2.5\,\mbox{k\ensuremath{\Omega}}$. Full
synchronization could be achieved with a narrower frequency range
$\Delta f_{self}$, a stronger coupling, or (as we will show) an altered
coupling topology. We chose this combination to highlight an asymmetry
in the synchronization that arises for positive $\alpha$: as coupling
decreases, the slowest oscillators remain intact while the faster
oscillators fall out of synchronization.

Due to this observed asymmetry, we hypothesized that we could increase
the degree of synchrony by silencing the slowest oscillators. By opening
the $S_{2}$ for some of the slowest oscillators, the equation describing
the dynamics changes in a simple but important way. The original Sakaguchi-Kuramoto
model includes a self-coupling term, made evident by pulling it out
of the all-to-all sum: 
\begin{equation}
\dot{\phi}_{i}=\omega_{i}-\frac{K\sin\alpha}{N}+\frac{K}{N}\sum_{j\ne i}\sin\left(\phi_{j}-\phi_{i}-\alpha\right).
\end{equation}
Opening $S_{2}$ effectively removes this self-coupling term for the
oscillator; from the standpoint of the oscillator just removed, we
have essentially increased its natural frequency by an amount $K\sin\alpha/N$.
From the standpoint of the other oscillators, the coupling has gotten
stronger, going from $K/N$ to $K/\left(N-1\right)$, and the width
of the population has gotten narrower because the lower end of the
flat spectrum has risen. Also, the self-coupled frequencies of the
oscillators at the higher end should be lower (due to higher self-coupling)
and the average velocity of the cluster should be closer to the high
end of the spectrum. Assuming that the oscillators are indexed in
order of increasing frequency and we have silenced the first $n$
of them, we have 
\begin{equation}
\dot{\phi}_{i}=\omega_{i}+\frac{K}{N-n}\sum_{j=n+1}^{N}\sin\left(\phi_{j}-\phi_{i}-\alpha\right).\label{eq:all-to-some}
\end{equation}
Thus the oscillators that are still ``talking'' should be more tightly
bound than they would have been with the other oscillator present,
while the just-removed oscillator will have a higher natural frequency
and be interacting with a stronger cluster. Silencing the slowest
oscillators should increase synchrony.

Figure \ref{ps}(b) reports a systematic test of this hypothesis.
There are N=19 oscillators altogether, and the horizontal axis records
how many of them are talking, i.e. $N-n$ (again, all are listening).
The vertical axis marks the value of $R_{B}$. Thus, Fig. \ref{ps}(b)
yields the behavior within accessible parameter space. The color indicates
the size of the largest cluster. Not surprisingly, as the coupling
is weakened, the synchronized state deteriorates. Full sync is achieved
for this system when the coupling is sufficiently strong and/or enough
of the slow oscillators have been silenced. Only toward the very right
of the figure do we start to lose the slower oscillators.

The figure also depicts the analytically predicted threshold for full
synchronization (line and filled circles). Here we first computed,
based on the Sakaguchi-Kuramoto model, the value of K at which all
the talkers fully synchronized, as well as the associated magnitudes
of the mean field $r$ and the coupled frequency $\Omega$. Then,
in a second step, it was determined if the non-talkers would be able
to entrain to that frequency for a driving signal with magnitude $r\cdot K$.
If not, we used an iterative approach to identify the minimum coupling
strength at which all oscillators synchronized. The iterative step
was only necessary for $N-n\le11$ and gives rise to the turning point
in the predictions. We see that the computed curve loosely follows
the experimental phase boundary.

The series presented in \ref{ps}(a) and \ref{ps}(c) correspond to
the letters in Fig. \ref{ps}(b). These series provide a detailed
snapshot of the behavior at different locations in parameter space.
When all oscillators are talking, we see that we lose oscillators
from the high-frequency side of the distribution exclusively, until
synchronization is completely destroyed. When only the five top oscillators
remain as talkers, shown in series D and E, global sync is achieved
at the largest coupling strengths, and as the coupling is weakened
the cluster loses oscillators on both sides of the distribution.

\begin{figure}
\includegraphics[width=3in]{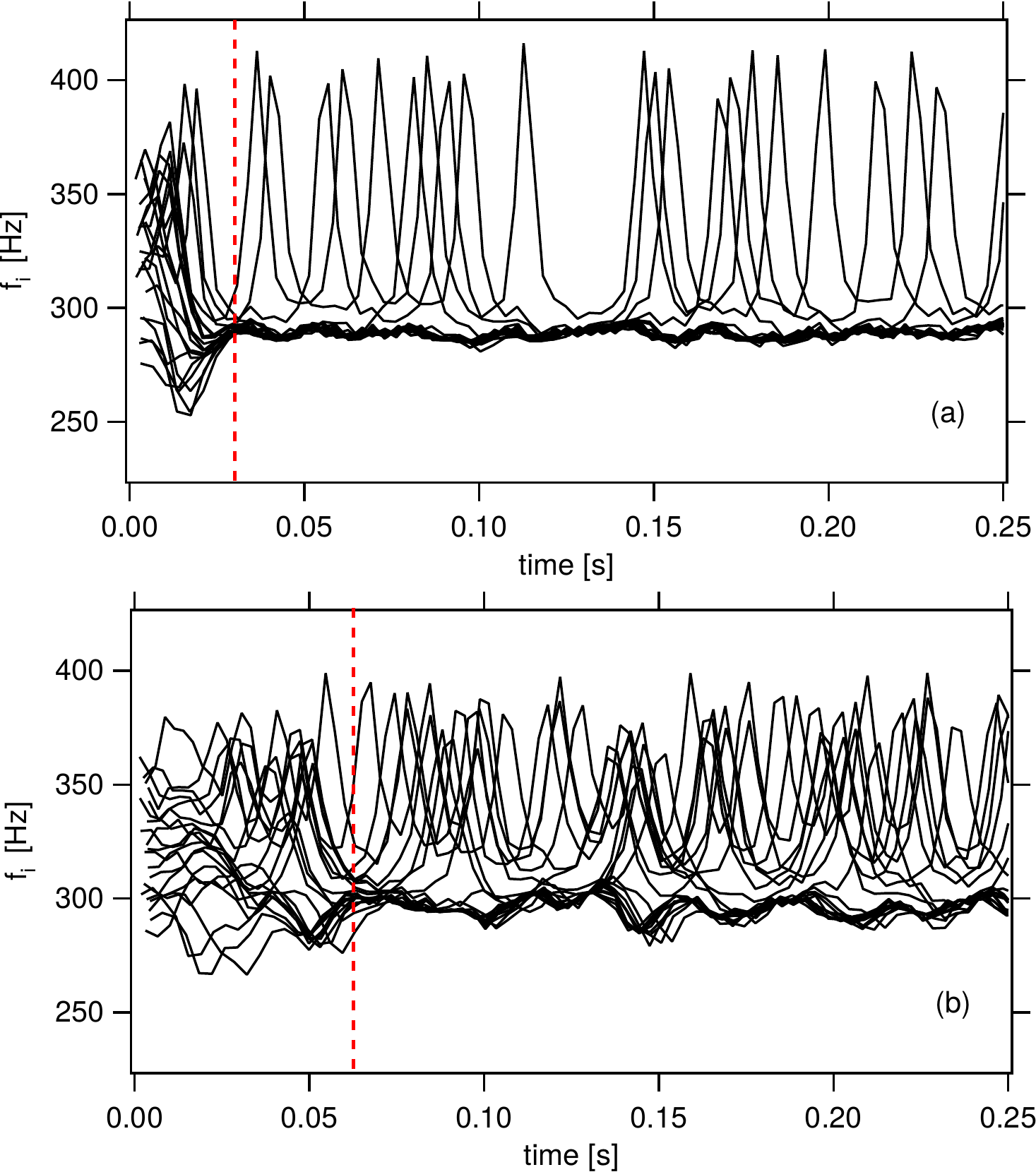} \caption{(Color online) Capturing the transients on route to synchronization.
Switches, $S_{1}$ and $S_{2}$, are closed for all oscillators. (a)
Full coupling strength ($R_{B}=4\,\mbox{k\ensuremath{\Omega}}$) and
(b) Reduced coupling strength ($R_{B}=3.5\,\mbox{k\ensuremath{\Omega}}$).
The dashed red line marks the approximate end of the transient and
the full formation of the synchronized cluster.}
\label{trans1} 
\end{figure}

Finally we turn to an experimental examination of the route towards
synchronization in this system. For this purpose, we capture the initial
transients before the steady state establishes itself. In Fig. \ref{trans1},
the dynamic frequency of each oscillator is displayed as a function
of time, where t=0 marks the time when the coupling is first turned
on. (Before t=0, the oscillators are allowed to free-run.) Here all
switches are closed and N=20. In Fig. \ref{trans1}(a), $R_{B}=4\,\mbox{k\ensuremath{\Omega}}$
and in (b) $R_{B}=3.5\,\mbox{k\ensuremath{\Omega}}$.

For the larger coupling strength, we observe that all but the fastest
oscillators manage to come in sync within a time of $\Delta t=0.03\,\mbox{s}$.
At $300\,\mbox{Hz}$, this corresponds to about 10 periods of oscillation.
When the coupling strength is lowered (Fig. \ref{trans1}(b)), the
transients last for a longer time, $\Delta t\cong0.06\,\mbox{s}$,
before the synchronized cluster has fully formed.

\begin{figure}[b]
\includegraphics[width=3in]{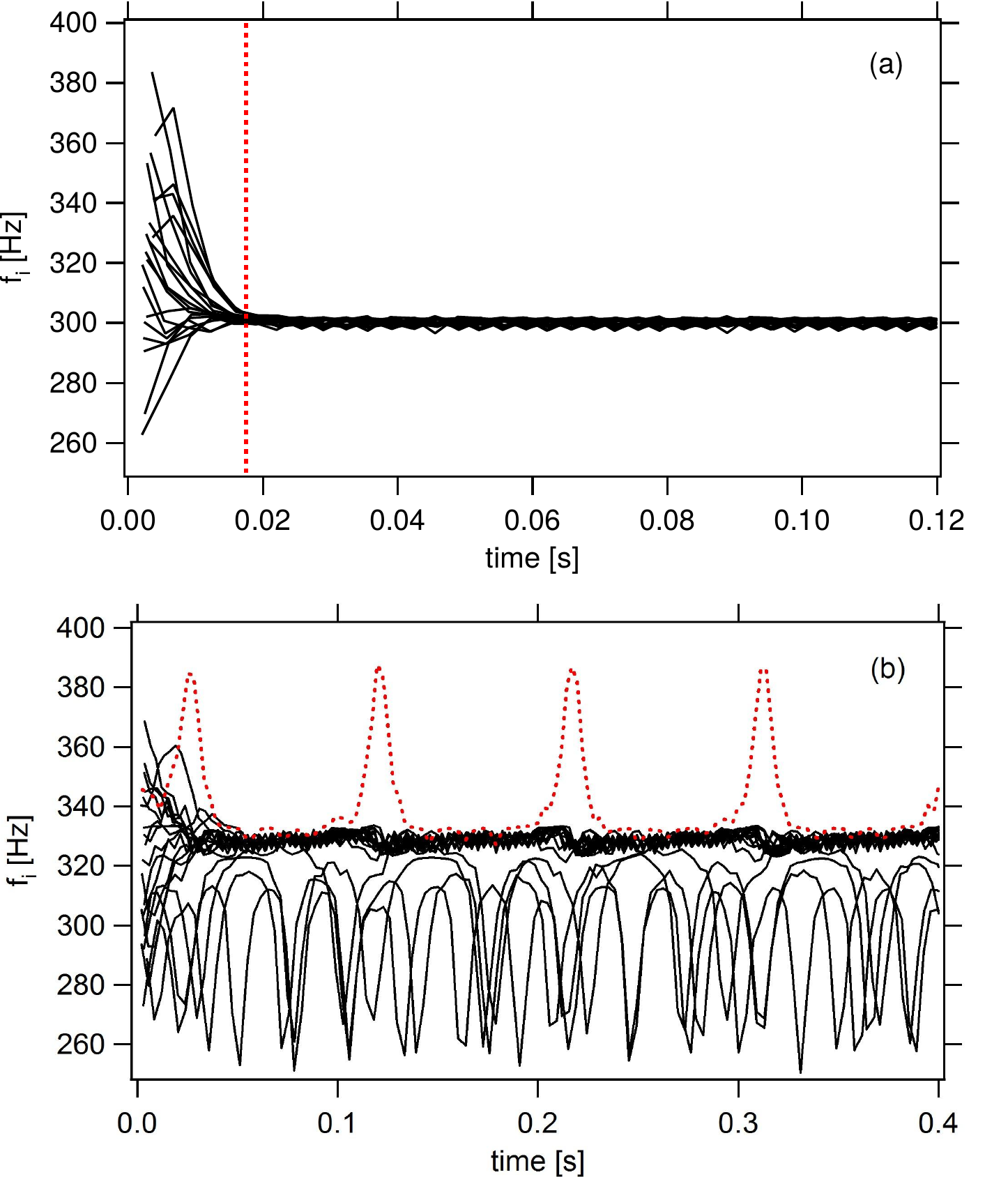} \caption{(Color online) Transients for the case where the switches S2 are open
for all but the top six oscillators. (a) Full coupling strength ($R_{B}=4\,\mbox{k\ensuremath{\Omega}}$)
and (b) Reduced coupling strength ($R_{B}=1.5\,\mbox{k\ensuremath{\Omega}}$).
Notice the oscillator with the highest natural frequency (depicted
in the dotted red line) experiences periodic bursts in frequency,
corresponding to times when its phase detaches from that of the cluster.
A few lower-frequency oscillators form a desynchronized random background.}
\label{trans2} 
\end{figure}

To illustrate the full range of dynamic behavior, it is instructive
to repeat this experiment for a point further to the right in parameter
space. For this purpose, we open switch $S_{2}$ for all but the fastest
six oscillators and consider two values of the coupling. Figure \ref{trans2}
depicts the results. At the largest coupling, Fig. \ref{trans2}(a)
clearly indicates global synchronization. In fact, all the oscillators
have been entrained within a time of $\Delta t<0.02\,\mbox{s}$, or
just a few periods. When the coupling is substantially lowered, a
number of interesting behaviors are observed, shown in see Fig. \ref{trans2}(b).
Here again the cluster is quickly formed, but we can clearly distinguish
the two kinds of unsynchronized behavior. One is a set of ``silenced''
oscillators, with lower frequencies, and the other is the top oscillator
which is also unable to follow the synchronized cluster. Both types
of unsynchronized oscillator exhibit the characteristic spike in dynamic
frequency indicating a rapid phase-slip event. However, the cluster
only responds, to the phase slip event of the faster, talking oscillator.
This is evident by the small dip in cluster frequency that slightly
lags the frequency spike of the fast oscillator. This lead/lag behavior
was also seen in the pairwise interactions (Fig. \ref{2oscdata}).
The magnitude of the change differs, reflecting a single oscillator
interacting with a cluster of many oscillators, in agreement with
the Sakaguchi-Kuramoto model.

\section{Conclusion}

In this work, we have studied a network of simple Wien-bridge oscillators
in an experimental regime for which they can be approximated as phase
oscillators. Using the experimental data for a pair of oscillators,
we determined the two adjustable parameters within the Sakaguchi-Kuramoto
model. We then tested detailed, quantitative predictions of this model
against experimental findings. Such a comparison yielded surprisingly
good agreement.

As an additional illustration of the effectiveness of the model, we
introduced and experimentally measured the behavior of a new kind
of coupling, termed all-to-some, in which all oscillators are coupled
to a signal that is constructed from only some of the oscillators.
For positive $\alpha$, we demonstrated that selective removal of
the slowest oscillators can improve the degree of global synchronization,
an observation that is consistent with the Sakaguchi-Kuramoto model.
The experimental data clearly shows the evolution towards collective
synchronization in this system when the coupling is turned on. It
also illustrates the way synchronization is lost when the coupling
is incrementally weakened.

These oscillators are ideal for synchronization experiments. They
are easy to build. As we have illustrated, the underlying phase dynamics
are well described by a simple model commonly studied in the synchronization
literature. The topology and nature of the interactions can be easily
prescribed. Nearly every aspect of the system can be easily measured.
In a field where numerical simulations are the norm, we believe that
a collection of electronic oscillators built using this design can
serve as a general-purpose system for real experimental tests of the
theories of synchronization.


\begin{thebibliography}{10}
\bibitem{buck} J. Buck and E. Buck, Sci. Am. \textbf{234}, 7485 (1976).

\bibitem{walker} T. J. Walker, Science \textbf{166}, 891 (1969).

\bibitem{kiss} I. Z. Kiss, Y. Zhai, and J. L. Hudson, Phys. Rev.
E \textbf{77}, 046204 (2008).

\bibitem{fukuda} H. Fukuda, N. Tamari, H. Morimura, and S. Kai, J.
Phys. Chem. A \textbf{109}, 11250 (2005).

\bibitem{solomon} M.S. Paoletti, C. R. Nugent, and T. H. Solomon,
Phys. Rev. Lett. \textbf{96}, 124101 (2006).

\bibitem{bridge} S. Strogatz, D. Abrams, A. McRobie, B. Eckhardt,
and E. Ott, Nature \textbf{438}, 43 (2005).

\bibitem{mertens} D. Mertens and R. Weaver, Phys. Rev. E \textbf{83},
046221 (2011).

\bibitem{zhang} M. Zhang, G. S. Wiederhecker, S. Manipatruni, A.
Barnard, P. McEuen, and M. Lipson, Phys. Rev. Lett. \textbf{109},
233906 (2012).

\bibitem{michaels} D. Michaels, E. Matyas, and J. Jalife, Circulation
Res. \textbf{61}, 704 (1987).

\bibitem{sompolinsky} H. Sompolinsky, D. Golomb, and D. Kleinfeld,
Phys. Rev. A \textbf{43}, 6990 (1991).

\bibitem{breakspear} M. Breakspear, S. Heitmann and A. Daffertshofer,
Front. Hum. Neurosci. \textbf{{4}}, 190 (2010).

\bibitem{kura} Kuramoto, Y., \emph{Chemical Oscillations, Waves and
Turbulence}. 1984, (Berlin: Springer Verlag).

\bibitem{sakaguchi} H. Sakaguchi and Y. Kuramoto, Prog. Theor. Phys.
\textbf{76}, 576 (1986).

\bibitem{kuramoto} Y. Kuramoto and I. Nishikawa, J. Stat. Phys.,
\textbf{49}, 569 (1987).

\bibitem{strogatz} S. H. Strogatz, Physica D \textbf{143}, 1 (2000).

\bibitem{pikovsky-book}A. Pikovsky, M. Rosenblum, J. Kurths, \emph{Synchronization:
A Universal Concept in Nonlinear Sciences}. 2003 (Cambridge university
press).

\bibitem{OttAntonsen}E. Ott T. M. Antonsen. Chaos \textbf{18}, 037113
(2008).

\bibitem{lasers}S. Yu. Kourtchatov, V. V. Likhanskii, A. P. Napartovich,
F. T. Arecchi, A. Lapucci, Phys. Rev. A \textbf{52}, 4089 (1995).

\bibitem{josephson}K. Wiesenfeld, P. Colet, S. H. Strogatz, Phys.
Rev. Lett. \textbf{76}, 404 (1996).

\bibitem{bz-reactors}J. Miyazaki, S. Kinoshita. Physical Review Letters
\textbf{96}, 194101 (2006).

\bibitem{bz-droplets}J. Delgado, N. Li, M. Leda, H. O. Gonzalez-Ochoa,
S. Fraden, I. R. Epstein. Soft Matter \textbf{7}, 3155-67 (2011).

\bibitem{bergner}A. Bergner, M. Frasca, G. Sciuto, A. Buscarino,
E. J. Ngamga, L. Fortuna, and J. Kurths. Physical Review E \textbf{85},
(2012): 026208.

\bibitem{FHN-circuit}L. V. Gambuzza, A. Buscarino, S. Chessari, L.
Fortuna, R. Meucci, M. Frasca. Physical Review E \textbf{90}, (2014):
032905.

\bibitem{rosenblum} A. A. Temirbayev, Z. Z. Zhanabaev, S. B. Tarasov,
V. I. Ponomarenko, M. Rosenblum, Phys. Rev. E \textbf{85}, 015204
(2012).

\bibitem{moro} S. Moro, Y. Nishio, and S. Mori, IEICE Trans. Fundamentals
\textbf{E78-A}, 244 (1995).

\bibitem{chain-universality}T. E. Lee, G. Refael, M. C. Cross, O.
Kogan, J. L. Rogers. Physical Review E \textbf{80}, 046210 (2009).

\bibitem{chimera-theory}D. M. Abrams, S. H. Strogatz. International
Journal of Bifurcation and Chaos \textbf{16}, 21 (2006).

\bibitem{reconstructing-networks}B. Kralemann, A. Pikovsky, M. Rosenblum.
Chaos \textbf{21}, 025104 (2011).

\bibitem{reconstruction}B. Kralemann, L. Cimponeriu, M. Rosenblum,
A. Pikovsky, R. Mrowka. Physical Review E \textbf{77}, 066205 (2008).\end{thebibliography}
\end{document}